\documentclass[12pt]{article}
\usepackage{amsmath}
\usepackage{graphicx}
\usepackage{amssymb}
\usepackage{amsfonts}
\usepackage{color}
\usepackage{indentfirst}
\usepackage{mathrsfs} 

\usepackage{soul}
\usepackage{cite}

\def\beq{\begin{eqnarray}}
\def\eeq{\end{eqnarray}}
\def\n{\label}                          
\newcommand{\eq}[1]{(\ref{#1})}

\def\ln{\,\mbox{ln}\,}

\def\al{\alpha}
\def\be{\beta}

\def\ga{\gamma}
\def\de{\delta}
\def\vp{\varepsilon}
\def\ep{\epsilon}

\def\ka{\kappa}
\def\la{\lambda}
\def\na{\nabla}

\def\si{\sigma}
\def\om{\omega}
\def\ph{\varphi}

\def\Ga{\Gamma}

\def\La{\Lambda}

\begin{document}

\begin{center}

\textbf{
\large Effective quantum gravity, cosmological constant
\\
and the Standard Model of particle physics}

\vskip 8mm

Breno~L.~Giacchini$^{(a)}$,
\ \ \
Tib\'erio~de~Paula~Netto$^{(a)}$,
\ \ \
Ilya~L.~Shapiro$^{(b)}$ \footnote{Also at Tomsk State
Pedagogical University.}
\vskip 8mm

(a) Department of Physics, Southern University of
Science and Technology,
\\
Shenzhen, 518055, China

(b) Departamento de F\'{\i}sica,  ICE,
Universidade Federal de Juiz de Fora,
\\
Juiz de Fora,  36036-900,  MG,  Brazil

\end{center}
\vskip 6mm

\begin{quotation}
\noindent
\textbf{Abstract.}
\ \
The renormalization group in effective quantum gravity can be
consistently formulated using the Vilkovisky and DeWitt version
of effective action and assuming a non-zero cosmological constant.
Taking into account that the vacuum counterpart of the cosmological
constant is dramatically different from the observed energy density
of vacuum, the running of the last quantity in the late cosmology
indicates strong constraints on the physics beyond the minimal
Standard Model of particle physics.
\end{quotation}

\section{\large Introduction}
\label{s1}

One of the main open issues in nowadays fundamental physics is
the origin of the so-called dark energy. The most likely candidate is
the cosmological constant because such a constant is an essential
element of the consistent quantum field theory (QFT) in curved space
(see e.g. \cite{OUP} for a recent discussion and further references).
On the other hand, there are many theoretically interesting alternative
models, assuming that the vacuum energy may vary with time (see, e.g,
\cite{Peebles2002,Copeland2006} as starting points). The question
is how one can expect to falsify these models using the existing
observational data. Indeed, the possibilities to do so are restricted to
the observation of the time-variable vacuum energy, or the effective
equation of state \cite{Sahni2002}.

Does the time dependence of the vacuum
energy mean that there is an essence more sophisticated than the
cosmological constant, governing the accelerated expansion of the
Universe? The answer is not obvious because the proper cosmological
constant may be changing for at least two reasons. The observable
cosmological constant is a sum of the two contributions, namely, the
fundamental constant in the action of gravity and the induced
counterpart owing to the phase transition and symmetry breaking
in the vacuum state \cite{Weinberg89} (see also \cite{CC-nova}
for the QFT aspects of the problem). The induced component could
have different magnitude because of the typical energy scale at
which the symmetry breaking occurs or, almost equivalently, to
restoration of the symmetry in the hot vacuum of the early Universe.
Furthermore, the observable cosmological constant may change
in the late Universe because of the renormalization group running.
The running of the vacuum component of the cosmological constant
in flat spacetime is a basic example of such a running (see e.g.
\cite{Brown}), that can be extended to semiclassical \cite{nelspan82}
and quantum \cite{frts82} gravity.

The running cosmology models are based on the universal form
of the scale-dependent density of the energy of vacuum, that can be
established using covariance arguments \cite{PoImpo} or the
assumption of a standard quadratic decoupling of massive degrees
of freedom in the IR, in the semiclassical approach
\cite{CC-nova,Babic},
\beq
\rho_\La = \rho^0_\La + \frac{3\nu}{8\pi G}\,
\big(H^2 - H^2_0 \big).
\label{runCC}
\eeq
Here  $\rho^0_\La$ and $H_0$ are the vacuum energy density and
the Hubble parameter at the reference point, e.g., in the present moment
of time. $\nu$ is a phenomenological parameter which cannot be
calculated with the known theoretical methods \cite{PoImpo,DCCrun}.
Since the particle creation from vacuum in the IR is suppressed
\cite{Opher-2004},
the energy conservation leads to the ``traditional'' logarithmic form of
the running for the Newton constant \cite{Gruni},
\beq
G(\mu)=\frac{G_0}{1+\nu\,\log\left(H^2/H_0^2\right)}\,.
\label{runGH}
\eeq
There is a possibility of a systematic scale-setting procedure
in the cosmological context \cite{Babic-setting}, providing the
identification of the Hubble parameter $H$ with the scaling
parameter $\mu$.

Let us stress that the aforementioned running in (\ref{runCC}) and
(\ref{runGH}) is owing to the quantum effects of \textit{massive}
particles in the IR, according to the corresponding decoupling
theorems \cite{AC,apco}. Usually, massless degrees of freedom do
not produce running of the dimensional parameters, such as $\rho_\La$
and $G$. On the other hand, (\ref{runGH}) looks like the one-loop
running of the dimensionless coupling in the Minimal Subtraction
scheme of renormalization, regardless it is derived in a very
different framework.

The models of running vacuum based on (\ref{runCC}) were
extensively explored and became an active field of research
(see e.g. \cite{PoImpo,DCCrun,Gruni,CC-pert,Jhonny-GrCo}
and further references in the last work). These models give a
well-motivated alternative to the theories of modified gravity
of all kinds, as in both cases one can describe a slowly
varying vacuum energy density.

From the theoretical side, one of the open questions is
whether the aforementioned universality of the running of vacuum
energy (\ref{runCC}) can be extended into the full quantum
gravity (QG). The definitive  answer to this question is unknown, as
there is no completely consistent theory of quantum gravity to
derive the running. On the other hand, any kind of a purely
metric quantum gravity should have massless and massive degrees
of freedom. The typical dimensional constant in quantum gravity
is the Planck mass $M_P$, therefore we can expect that all massive modes
have the mass of this magnitude \cite{seesaw}. For the
cosmological applications, what we need is an effective formulation
of QG, when the massive modes are assumed to decouple (see, e.g.,
\cite{don-reviews,Burgess} and references therein). In this regime,
we expect to meet the universal IR theory of QG based on the quantum
general relativity (GR).
Therefore, we need to account only for the massless modes
\cite{DalMaz,effQG} and it seems this can be a solid basis for
deriving the QG-based running of $\rho_\La$ and $G$ that can be
applied in the late running-vacuum cosmology.

Now we are coming close to the main subject of this contribution,
i.e., the role of the ``unique'' effective action of Vilkovisky and DeWitt
for the IR running in the effective QG. In the framework of usual
perturbative field theory, the running of $\rho_\La$ and $G$ is
possible only if the initial gravity theory has a non-zero cosmological
constant. However, even in this case, the individual running of both
quantities depends on the gauge fixing and on the parametrization
of the quantum metric. Only the running of the dimensionless ratio of
these parameters is universal \cite{frts82}, but this is insufficient
for the cosmological applications. The construction of
Vilkovisky--DeWitt effective action resolves this difficulty at the
one-loop \cite{Vilk-Uni} and higher-loop \cite{DeWitt-ea} levels.
This useful feature of the ``unique'' effective action is owing to
its geometric, covariant formulation in the space of the fields.
The renormalization group running of  $\rho_\La$ and $G$, based on
Vilkovisky's construction has been explored in \cite{TV90}. In
the recent works \cite{UEA-contas,UEA-RG} we performed an explicit
verification of the gauge- and
parametrization-invariance of these
renormalization group equations and proved that, in the effective QG
framework, these equations are exact, i.e., not restricted by the
one-loop approximation, as explained also in Sec. \ref{s3} below.

The status of the Vilkovisky--DeWitt effective action has been
previously discussed in the work \cite{KKT-88} based on three
interesting applications where the gauge fixing and/or parametrization
ambiguities in the usual effective action do not enable one to achieve
the desired qualitative output, while the geometric methods provide the
result. The general conclusion of this consideration is that the
Vilkovisky--DeWitt construction is not a panacea, nor a placebo,
but rather a useful tool for making calculations.
In the present contribution, we discuss the application of the
effective QG-based running \cite{UEA-RG} of $\rho_\La$ and $G$
in cosmology and show that these effects may provide dramatic
consequences for particle physics. This can be viewed as one more
application where the  Vilkovisky--DeWitt geometric approach
yields a nontrivial consequence.
Since the technical details of the Vilkovisky formalism are not
the main topic of this paper, we postpone to the
Appendix~\ref{Appendix} a brief
review of it and a discussion about its possible limitations.
However, we would like to stress, from the very beginning,
that this approach is an additional independent input and that its
consequences may be verified or falsified only by means of
experiments or observations.

The manuscript is organized as follows.
In Sec. \ref{s2}
we briefly review the cosmological constant and the fine-tuning
problem, from the standard perspective of \cite{Weinberg89} and
\cite{CC-nova}.
Sec. \ref{s3} discusses the exact effective QG running
\cite{UEA-RG} and explains why it might break down the
fine-tuning for the cosmological constant.
In Sec. \ref{s4} we perform the numerical estimates of the
mentioned breaking for a few models of particle physics
and arrive at severe constraints on the physics beyond the
standard model. Finally, in Sec. \ref{s5} we draw our
conclusions and add some extra discussion.
%

\section{\large Cosmological constant problem and the running}
\label{s2}

As it was already mentioned above, the running  of $\rho_\La$ and
$G$ in GR-based QG is possible only because of the
non-zero cosmological constant. The higher loop  contributions
come with growing powers of the dimensionless ratio
$\rho_\La G^2 = \rho_\La\,M_P^{-4}$. It is important to stress
that $\rho_\La$ in this expression is \textit{not} the observable
vacuum energy density $\rho^{obs}_\La$. It is well-known
\cite{Weinberg89} (see also \cite{CC-nova}) that
\beq
\rho^{obs}_\La\,=\, \rho_\La + \rho^{ind}_\La \,,
\label{CCobs}
\eeq
where $\rho_\La$ and $\rho^{ind}_\La$ are the vacuum and the induced
densities of cosmological constant, respectively. An independent
quantity $\rho_\La$ is a necessary element of the renormalizable
semiclassical theory
(see, e.g., \cite{OUP} for an introduction). Loop corrections without
external lines of matter fields produce divergences, including those
without derivatives of the metric tensor. These divergences require
renormalization and, in particular, fixing the renormalization
condition. As usual with independent parameters, this procedure
involves a measurement. Thus, the value of $\rho_\La$ can be
defined only from the cosmological observations of $\rho^{obs}_\La$.
After that, one has to subtract $\rho^{ind}_\La$ to arrive at the
value of $\rho_\La=\rho^{vac}_\La$.

From the theoretical side, the minimal magnitude of $\rho_\La$ is
defined by its running in a semiclassical theory. For example, in the
minimal standard model (MSM) this indicates at a value of the order
of the fourth power of the Fermi mass, $M_F^4$. As $\rho_\La$ is an
independent parameter, its value at the reference scale $\mu_0$ can
be defined only from the experimental or observational data. On the
other hand, $\rho^{ind}_\La$ is, in principle, calculable from the
underlying matter fields model. If its origin is the spontaneous
symmetry breaking (SSB) in the MSM, we have the well-known
relation with the VeV of the Higgs field,
$\rho^{ind}_\La \sim \la v^4 \approx 10^8 \, \text{GeV}^4$.
As the value of $v$ is defined by the typical (Fermi) energy scale
$M_F \approx 293\, \text{GeV}$, in what follows we
shall associate
$\rho^{ind}_\La$ with $M_F^4$. In case there is another, similar,
phase transition at higher energy scale such as $M_X$, we have
to replace $M_F$ by $M_X$ in both $\rho_\La$ and $\rho^{ind}_\La$.

It is remarkable that the theoretical predictions for the two
ingredients in the \textit{r.h.s.}~of~(\ref{CCobs}) give the same
order of magnitude. At the same time, the relation (\ref{CCobs})
is famous for the huge amount of fine-tuning required for the
cancelation in its \textit{r.h.s.}, providing a very small value
for the observable sum. In the particle physics units, the value
of $\rho^{obs}_\La$,  it is about  $10^{-47}\, \text{GeV}^4$, such that
even in the MSM we need about 56 orders of the fine-tuning in the
choice of the renormalization condition $\rho_\La(\mu_0)$ .

The cosmological constant problem is a real mystery, as the 56-order
fine tuning can be violated even by very small changes in the Yukawa
couplings that enter the game via the one-loop or higher-loop corrections
(up to 21 loops). It is worth noting that even a small mismatch in
the choice of the condition for $\rho_\La(\mu_0)$ may lead to either
a negative, zero or too big (more than 100 times greater) positive value
of  $\rho_\La^{obs}$. All  three options contradict our own existence
 through the anthropic arguments \cite{Weinberg-87antrop}. In the next
sections we shall see that the well-defined running $\rho_\La(\mu)$
in effective QG imposes strong constraints on the particle physics
beyond MSM.

\section{\large On the running in effective QG}
\label{s3}

The starting point in the discussion of the running in effective QG is
the gravitational action,
\beq
S \,=\, -\,\frac{1}{16\pi G}\int d^4x \sqrt{-g} \,(R + 2\La),
\label{EH}
\eeq
where $\rho_\La = \frac{\La}{8\pi G}$,
together with the theory of gauge-invariant renormalization
(see e.g. \cite{OUP} and references therein) and the power
counting formula for quantum GR,
\beq
\om(\mathcal{G}) + d(\mathcal{G}) = 2 + 2L - 2K_\La.
\label{power}
\eeq
Here $L$ is the number of loops in the given diagram
$\mathcal{G}$ and $K_\La$ is the number of vertices coming from
the cosmological  constant term. For the logarithmic divergences
$\om(\mathcal{G})$ is zero and then the last formula gives the
number of derivatives $d(\mathcal{G})$ in the corresponding
counterterm.

It is easy to see that for $K_\La=0$ we never get the renormalization
of the Einstein--Hilbert term, but only higher-derivative counterterms.
However, the situation is different in the case $K_\La \neq 0$. One of
the main observations of \cite{UEA-RG} and of the present work is
that $\La$ in (\ref{EH}) does not correspond to $\rho^{obs}_\La$ but,
instead, to the $\rho_\La = \frac{\La}{8\pi G}$ in the vacuum part of
(\ref{CCobs}). As we have seen in the previous section, the two
quantities $\rho^{obs}_\La$ and $\rho_\La$  are dramatically different
and this may change the game in the effective QG-based running.

The important aspect in the effective QG is whether we are capable
of obtaining results which are free of ambiguities. For instance, the
known theorems about gauge-fixing and parametrization dependence
\cite{aref,volatyu} tell us that the one-loop divergences
of the effective action,
\beq
\Ga^{(1)}_{div} =\frac{1}{\ep}\,
\int d^4x\sqrt{-g}\,\big\{
c_1\,R_{\mu\nu\al\be}^2 + c_2R_{\al\be}^2 + c_3R^2
+ c_4  {\Box}R + c_5R + c_6 \big\},
\label{Gamma1}
\eeq
are universal only on the classical mass shell. Here
$\ep=(4\pi)^2(n-4)$ is the parameter of dimensional regularization.
Using the approach of \cite{frts82,a} we arrive at the relation
\beq
&&
\Ga^{(1)}_{div}(\al_i) \,-\, \Ga^{(1)}_{div}(\al^{0}_i)
\,\,=\, \frac{1}{\ep}\,
\int d^4x\sqrt{-g}\,\Big(
b_1 R_{\mu\nu} + b_2 Rg_{\mu\nu}
 \qquad
 \qquad
\nonumber
\\
&&
 \qquad
 \qquad
 \qquad
 \qquad
 \qquad
 + \,\,b_3 g_{\mu\nu} \La
 + b_4 g_{\mu\nu}\Box + b_5 \na_\mu \na_\nu\Big)\,\vp^{\mu\nu}\,,
\nonumber
\eeq
where
$b_k=b_k(\al_i)$ and $\al_i$ represent the full set of parameters
defining an arbitrary gauge-fixing and parametrization of the metric,
and $\vp^{\mu\nu}= G^{\mu\nu} - \La g^{\mu\nu}$ are the classical
equations of motion. The two invariant quantities are
\begin{equation}
c_1 \qquad \text{and}
\qquad
c_{\rm inv}\,=\,c_6 - 4\La c_5 + 4\La^2 c_2 + 16\La^2 c_3,
\label{invs}
\end{equation}
which means that the invariant running is possible only for the
dimensionless combination of $G$ and $\La$ \cite{frts82,JDG-QG}.

The Vilkovisky \cite{Vilk-Uni} and DeWitt \cite{DeWitt-ea}
definition of effective action is different, as it is based on the
covariant calculus in the space of physical fields.
A brief review of the formalism can be found in the Appendix~\ref{Appendix},
where we also discuss some subtleties involving its application
to QG. More detailed considerations can be found, e.g., in
\cite{bavi85,Huggins:1987zw,TV90,UEA-RG,UEA-contas,Fradkin:1983nw,Rebhan:1987cd,Ellicott:1987ir,KKT-88}.

The renormalization group running of $G$ and $\La$, in the effective
QG based on the Vilkovisky--DeWitt version of effective action, has
the form \cite{TV90,UEA-RG}
\beq
G(\mu)
\,=\,
G_0\,\Big[1+ \frac{10}{(4\pi)^2}\,\ga_0\ln \frac{\mu}{\mu_0}\Big]^{-4/5}
\label{sol-ka}
\eeq
and
\beq
\La(\mu)
\,=\,
\La_0\,
\Big[1+ \frac{10}{(4\pi)^2}\,\ga_0\ln \frac{\mu}{\mu_0}\Big]^{-1/5},
\label{sol-La}
\eeq
where $\ga = 16\pi G\,\La$ and the subscript label zero indicated
the value $\mu_0$
of the reference scale, e.g., the present epoch of the Universe. Thus,
$\ga_0 = 16\pi G_0\,\La_0 = 128\pi^2 G_0^ 2\rho_\La^0
= 128\pi^2 \rho^0_{\La} M_P^{-4}$.

Let us stress that, differently from (\ref{runCC}) and (\ref{runGH}),
Eqs.  (\ref{sol-ka}) and (\ref{sol-La}) come from a real calculation,
based solely on the assumption of applicability of the Vilkovisky
and DeWitt effective action in quantum field theory.

According to the power counting (\ref{power}), the higher loop
corrections to these equations are proportional to higher powers
of $\ga$ and, for a sufficiently small $\La$, can be regarded
as negligible.
Starting from this point and taking into account the arguments from
Sec. \ref{s2}, we can explore the physical consequences of the
running  (\ref{sol-ka}) and  (\ref{sol-La}).

\section{\large Cosmology with running vacuum in effective QG}
\label{s4}

To evaluate $\rho_\La$ and $\ga_0$ we have to remember that
$\rho_\La$ has approximately---and with a great precision---the
same absolute value as $\rho^{ind}_\La$.
It is clear that $\rho_\La \gg \rho^{obs}_\La$,
but numerically $\ga_0$ is still a small quantity. For instance, in
the theory when the MSM is valid until the Planck energy scale,
we have $\ga_0 \sim 10^{-65}$, while for the SUSY GUT this
coefficient may be $\ga_0 \sim 10^{-12}$--$10^{-8}$. Do these
small numbers guarantee that the variations caused by (\ref{sol-ka})
and  (\ref{sol-La}) are irrelevant?

To evaluate the consequences of the running, let us consider
the strongest option, that is, the SUSY GUT case.
Then, the value of $\rho_\La$ should be of the order of $M_X^4$,
assuming $M_X \sim 10^{16}$~GeV. Accordingly, the parameter
$\ga$ is of the order of $(M_X/M_P)^4$. In this case, we find the
strong inequality $\frac{10\ga_0}{(4\pi)^2}\ll 1$. After a small
algebra, (\ref{sol-ka}) and  (\ref{sol-La}) boil down to
\beq
G(\mu)
\,=\,
G_0 \Big[1\,-\, \frac{8}{(4\pi)^2}\,\ga_0\ln \frac{\mu}{\mu_0}\Big]
\label{sol-ka-1}
\eeq
and
\beq
\rho_\La(\mu)
\,=\,
\rho^0_{\La}\,
\Big[1 + \frac{6}{(4\pi)^2}\,\ga_0\ln \frac{\mu}{\mu_0}\Big].
\label{sol-La-1}
\eeq
The derivation of the last equation requires
expanding the right-hand side of both (\ref{sol-ka}) and  (\ref{sol-La})
up to the first order in the small parameter $\ga_0$ and replacing
the result into the formula $\rho_\La = \frac{\La}{8\pi G}$.

Remember that the standard identification of scale in cosmology
is $\mu\propto H$ \cite{CC-nova,PoImpo,Babic-setting}. Taking
this identification,
the effects of (\ref{sol-ka-1}) and  (\ref{sol-La-1})
are dramatically different. Indeed, while for $G(\mu/\mu_0)=G(H/H_0)$
there is a usual slow logarithmic running that is not too relevant in
cosmology,
the effect of the cosmological constant running
(\ref{sol-La-1}) may be strong.

In the SUSY GUT case,
a simple calculation gives
\beq
\frac{6}{(4\pi)^2}\,\ga_0 \,\sim\, 48 \Big(\frac{M_X}{M_P}\Big)^4
\,\approx\, 10^{-11}.
\label{susygut-ga}
\eeq
This is a really huge number, because it has to be multiplied
not only by the logarithmic factor but, at the first place, by the
$\rho^0_{\La}$, i.e., by the \textit{vacuum} energy density in
the \textit{r.h.s.} of the main relation (\ref{CCobs}). In SUSY
GUT the value
of $\rho^0_{\La}$ is about 111 orders of magnitude
greater than the observed value $\rho^{obs}_{\La}$. Thus,
the running (\ref{susygut-ga})
produces a discrepancy with the cosmological observations
proportional to $10^{100}$ (googol)
for a change of about one order
of magnitude in the parameter $H$. Needless to say that this result
contradicts the anthropic calculations \cite{Weinberg-87antrop}.
Thus, we have to give up either on the effective QG based on
the Vilkovisky ``unique'' effective action, or on the SUSY GUT
and the corresponding generation of induced vacuum energy in
the \textit{r.h.s.} of (\ref{CCobs}). As the present report is
devoted to the effective QG-based running, we conclude that
the SUSY GUT hypothesis fails in this framework.

Let us consider another extreme of the energy scale and assume
that the MSM is valid up to the Planck scale.
In this case, instead of (\ref{susygut-ga}) we meet
\beq
\frac{6}{(4\pi)^2}\,\ga_0 \,\sim\, 48 \Big(\frac{M_F}{M_P}\Big)^4
\,\approx\, 10^{-65}.
\label{MSM-ga}
\eeq
When multiplied by $\rho^0_{\La} \sim M_F^4$, we find the
variation of the observed cosmological constant given by \
$\de \rho^{obs}_{\La} \approx 10^{-55} \ln (H/H_0) \,\text{GeV}^4$.

Taking the range of change of $H$ between the inflationary epoch with
$H_{infl} \leqslant 10^{15}\,\text{GeV}$ and the present-day Universe
with $H_0 \approx 10^{-42}\,\text{GeV}$, the logarithmic factor is
$\ln (H/H_0) \approx 131$. Then the numerical estimate
based on (\ref{MSM-ga}) gives
$\de \rho^{obs}_{\La} \approx 10^{-53} \, \text{GeV}^4$,
 that is just
six orders of magnitude smaller than the observed value
$\rho^{obs}_{\La}$. At this point we can make two observations:

1. Our model of the effective QG running is lucky enough to pass
the test related to MSM. This means, e.g., that the experimentally
confirmed model of particle physics does not contradict the
anthropic arguments. The opposite output would mean the
disproval of the Vilkovisky and DeWitt approach.

2. Since the result is proportional to $M_F^8$, we can state that
the existence of new physics based on the symmetry breaking
beyond the energy scale of $10M_F$ contradicts the effective
running of the cosmological constant. It is remarkable that the
effective QG provides such a relation between the cosmological
constant problem and the particle physics.

\section{\large Conclusions and discussions}
\label{s5}

The running derived in the effective QG based on the
Vilkovisky--DeWitt effective action enables one to formulate the
link between particle physics and cosmology. In particular, we
find that the MSM with the corresponding SSB  leads to the running
of the vacuum cosmological constant
that does not contradict the cosmological observations and, in
particular, the anthropic restrictions derived by Weinberg
\cite{Weinberg-87antrop}.

On the other hand, the mentioned running of the cosmological
constant imposes severe restrictions on the SSB and the generation
of induced cosmological constant in
the physics beyond the MSM. Even assuming the symmetry
breaking at the $10\,\text{TeV}$ scale means we may run out of the
scope with the cosmological constant violating the fine tuning in
(\ref{CCobs}). The energy scale below $10\,\text{TeV}$ is explored
in LHC, but this does not mean that new physics beyond MSM is
``forbidden'' by the effective QG and the corresponding running.
The obtained restrictions leave a lot of space for constructing
particle physics models beyond MSM. Thus, it would be
interesting to explore this possibility in more details.
The corresponding analysis is beyond the scope
of the present work and will be presented as a separate publication.
Let us just note that the limitation concerns the value of the new
Higgs-like vacuum expectation value (VeV) and not the masses
of the particles. Anyway, the preliminary result is that many
(albeit not all) GUT models and supersymmetric extensions of
the MSM may be ruled out by the new criterion based on quantum
gravity. On the other hand, there are theories, e.g., based on the
technicolor approach, which may have rather large masses of
the particles beyond the standard model and still escape the
restrictions discussed in the present work.

Finally, the well-defined gauge and parametrization independent
running (\ref{runCC}),
originally described by Taylor and Veneziano \cite{TV90} and
explored in the effective framework in \cite{UEA-RG}, provides
interesting hints concerning the connection between different
branches of Physics and also opens new horizons for further
work.

\appendix
\section{\large Brief review and discussion of the Vilkovisky effective action for QG}
\label{Appendix}

In this appendix we make a short presentation of the Vilkovisky
effective action and comment on the possible limitations of the
formalism, with a focus on the choice of metric in the space
of fields in QG. This discussion is done in more detail than in
our previous publications \cite{UEA-contas,UEA-RG}.

The Lagrangian quantization of gauge theories, including QG, involves
fixing a gauge which breaks the classical action's gauge invariance.
In DeWitt's background field method (see~\cite{Abbot82} for review
and references) the resultant effective action is
covariant with respect to gauge transformations of the background field,
but it still depends on the choice of gauge fixing for the quantum field.
The difference between these two situations is sometimes referred to as
``gauge invariance'' and ``gauge-fixing dependence''~\cite{KKT-88}.

In quantum GR, the background field method allows the evaluation
of the divergent part of the effective action. Owing to locality,
the divergences are scalars constructed with the curvature tensors and
their covariant derivatives [see Eq.~(\ref{Gamma1})].
As discussed in Sec.~\ref{s3}, the expression~(\ref{Gamma1}) is
\textit{invariant} under spacetime diffeomorphisms, although some of
the coefficients $c_i$ depend on the gauge chosen for the quantized
field, i.e., they are \textit{gauge-fixing dependent}.

The gauge-fixing dependence of the effective action can be
regarded as part of the more fundamental dependence on the field
parametrization, which also affects non-gauge theories~\cite{Vilk-Uni,volatyu}.
At one-loop level this can be understood by recalling that
the Hessian of the classical action, $\frac{\de^2 S}{\de \ph^i \de \ph^j}$,
does not behave as a tensor under redefinitions of the field $\ph^i$.
In the same spirit as in Riemannian geometry, Vilkovisky~\cite{Vilk-Uni}
introduced an affine structure on the configuration space
$\mathscr{M}_{\perp}$ of physical fields, with a metric $\bar{G}_{ij}$ and a
connection $\mathscr{T}^k_{ij}$, and modified the definition
of the effective action such that it transforms in a covariant manner
under diffeomorphism, namely,
\beq
\exp i \Ga(\ph)
\,=\,
\int \mathcal{D} \ph' \mu(\ph^{\prime})
\,\exp\left\lbrace
i \left[ S(\ph^{\prime}) + \si^i(\ph,\ph^{\prime}) \Ga_{,i}(\ph) \right]
\right\rbrace,
\label{UEA_def}
\eeq
where $\mu(\ph^{\prime})$ is an invariant functional measure and
$\si_i(\ph,\ph^\prime)$ is the derivative with respect to $\ph^i$ of
the world function $\si(\ph,\ph^{\prime})$~\cite{BDW-65,J.L.Synge:1960zz}. Note that
the affine structure is defined on the space of physical fields,
which means that if $\mathscr{G}$ is a gauge group acting on a
space $\mathscr{M}$ of fields $\ph^i$, then
$\mathscr{M}_{\perp} = \mathscr{M} / \mathscr{G}$. Therefore,
since $\si^i(\ph,\ph^\prime)$ behaves as a vector with respect to
$\ph^{i}$ and as a scalar with regard to $\ph^{\prime i}$, the effective
action $\Ga (\ph)$ defined by~\eqref{UEA_def} is gauge invariant and it
is independent of the parametrization and gauge fixing choices.
Because of this, the covariant effective action $\Ga(\ph)$ is
also known as ``unique effective action''.

The metric $G_{ij}$ in the full configuration space $\mathscr{M}$ of
fields is obtained through the following criteria: (i) It should be
an ultra-local quantity and do not contain derivatives of the fields
in order to not violate the $S$-matrix theory. (ii) For quadratic
non-interacting field theories, $G_{ij}$ should provide a flat field
space.  (iii)  It must be uniquely determined by the classical
action $S(\ph)$ of the theory, namely, it should be chosen as the
local metric contained in the highest-derivative term of the classical
action after projecting out the gauge-dependent degrees of freedom.
This is known as Vilkovisky's prescription for the choice of metric
in the space of fields~\cite{Vilk-Uni}.

The projection onto the space $\mathscr{M}_{\perp}$ is performed
by the operator~\cite{Vilk-Uni,Huggins:1987zw,Fradkin:1983nw}
\beq
\Pi^{i} {}_{j} = \de^{i} {}_{j} - R^{i}_\al (N^{-1})^{\al\be} R^{k}_\be G_{kj},
\eeq
where $R^{i}_\al$ are the generators of gauge transformations
and $(N^{-1})^{\al\be}$ is the inverse of the metric $N_{\al\be}$
on the gauge group,
\beq
N_{\al\be} = R^{i}_\al G_{ij} R^{j}_\be.
\eeq
Therefore, the projection of the metric on $\mathscr{M}_{\perp}$ is
\beq
\bar{G}_{ij}
\,=\,
\Pi^k {}_i  G_{k\ell} \Pi^\ell {}_j
\, = \, G_{ij} - G_{ik} R^k_\al (N^{-1})^{\al\be} R^\ell_\be G_{\ell j}.
\eeq
The affine connection $\mathscr{T}^k_{ij}$ can then be obtained by
requiring its compatibility with the physical field-space metric,
$\bar{\na}_k \bar{G}_{ij} = 0$ (see, e.g.,
\cite{Vilk-Uni,Kunstatter:1986qa,Huggins:1987zw} for further
comments and explicit formulas).

Notice, however, that by projecting out the gauge-dependent
part of the field we obtain
\beq
\n{red}
\ph^i_\perp \equiv \Pi^i {}_j \ph^j
= \ph^{i} - R^{i}_\al (N^{-1})^{\al\be} R^{k}_\be G_{kj} \ph^j.
\eeq
Thus, in the so-called Landau--DeWitt gauge, defined by
\beq
\n{4}
R^{k}_\be G_{kj} \ph^j = 0 ,
\eeq
we have
\beq
\n{blue}
\ph^i_\perp = \ph^{i}.
\eeq
This means that Vilkovisky's prescription for $G_{ij}$ is equivalent
to getting the metric which follows from the theory's action using the
Landau--DeWitt gauge~\cite{Fradkin:1983nw}. We stress that this is only
due to the fact that in the gauge~\eqref{4} the identity~\eqref{blue} holds,
and that at this stage one is not choosing a specific gauge which could,
potentially, simplify calculations. Thus, the aforementioned detail
does not reduce the generality of the scheme.

Let us discuss in more detail the choice of the
metric $G_{ij}$ in the case of QG.
In metric theories of gravity, $\ph^i = g_{\mu\nu}$ and
there is a one-parameter family of metrics on $\mathscr{M}$ in
accordance with (i)-(ii),
\beq
\n{sm}
G_{ij} = G^{\mu\nu,\al\be} = \tfrac12 ( \de^{\mu\nu,\al\be}
+ a g^{\mu\nu} g^{\al\be} ),
\qquad
\de^{\mu\nu,\al\be} = \tfrac12 ( g^{\mu\al} g^{\nu\be}
+ g^{\mu\be} g^{\nu\al} ) ,
\eeq
where $a\neq -1/4$ is a constant. The value $a = -1/4$ is discarded,
as otherwise the metric~\eq{sm} has no inverse.
In order to fix the parameter
$a$, we shall follow Vilkovisky's prescription.

Particularizing the discussion for quantum GR, the expansion
of the Einstein--Hilbert action (\ref{EH}) in the background
field method splitting 
\beq
\n{bfm}
g_{\mu\nu} \,\,\longrightarrow \,\, g'_{\mu\nu}\,=\,
g_{\mu\nu} + \ka h_{\mu\nu}, \qquad \ka^2 = 16\pi G,
\label{baparam}
\eeq
where $g_{\mu\nu}$ is the background field and $h_{\mu\nu}$ is
the quantum field, gives the quadratic form
\begin{equation}
\begin{split}
\n{bili}
S^{(2)} = & - \frac12 \int d^4 x \sqrt{-g} \big\{ h_{\mu\nu}
\left[ \tfrac12 ( \de^{\mu\nu,\al\be} - \tfrac12 g^{\mu\nu} g^{\al\be} ) \Box
\right] h_{\al\be}
\\
& + (\na_\la h^{\mu\la} - \tfrac12 \na^\mu h)^2 \big\} + \cdots .
\end{split}
\end{equation}
Here the ellipsis stands for terms without derivatives of the quantum
field, which are unimportant for the present discussion. Since the
generators of the spacetime diffeomorphisms are
\beq
R^{i}_\al \equiv R_{\mu\nu,\al}
= - (g_{\mu\al} \na_\nu + g_{\nu\al} \na_\mu),
\eeq
a simple calculation shows that the Landau--DeWitt gauge
condition~\eqref{4} reads
\beq
R^{k}_\al G_{kj} \ph^j = \na_\mu h^{\mu}_\al + a \na_\al h = 0
\quad
\Longrightarrow \quad \na_\mu h^{\al\mu} = - a \na^\al h.
\eeq
Substituting this condition into~\eq{bili} and recalling~\eqref{blue},
we find the bilinear projected action on $\mathscr{M}_{\perp}$,
\beq
&&
\quad
S^{(2)}_\perp
\,=\,
- \frac12 \int d^4 x \sqrt{-g}\,  h_{\mu\nu}^\perp \Big\{
\frac12 \big[
\de^{\mu\nu,\al\be} - \big(2 a^2 + 2 a + 1\big)
g^{\mu\nu} g^{\al\be} \big]\,\Box
\Big\}  h_{\al\be}^\perp \,+\, \cdots.
\eeq
Thus, the requirement that the term between curly brackets in
the last expression equals the metric \eq{sm} leads to the algebraic
equation~\cite{Vilk-Uni,Fradkin:1983nw}
\beq
\n{eqparaa}
- (2 a^2 + 2 a + 1) \,=\, a,
\eeq
which has the solutions
\beq
a = - \frac12 \qquad \text{and} \qquad a = - 1.
\eeq
The value $a = - 1$ is discarded because the metric on the
gauge group
\beq
N_{\al\be} = R^{i}_\al G_{ij} R^{j}_\be = - \big[
g_{\al\be} \Box + (1 + 2 a) \na_\al \na_\be + R_{\al\be}\big]
\eeq
becomes degenerate. Thus, for the QG based on GR,
in the simplest parametrization (\ref{baparam}), the Vilkovisky's
prescription for the choice of metric gives $a = - 1/2$ in an
unambiguous way.

Furthermore, since the metric on $\mathscr{M}$ transforms as a
tensor under field reparametrizations, Eq.~\eq{eqparaa}
holds even for quantum metric parametrizations more general
than \eq{bfm}. Thus, the ambiguity represented
by the coefficient $a$ is fixed in the same way, as explicitly shown
in~\cite{UEA-contas}. In this paper (as well as in our
work~\cite{UEA-RG}) we assume this choice of the metric.

This elementary exposition of the Vilkovisky formalism enables us
to discuss some of the usual criticism about it.
First of all, we notice that if the conditions (i) and (ii) for the field-space
metric are satisfied, the Vilkovisky effective action produces the same
result as the ordinary definition of the effective action at on-shell level.
In particular, both formalisms generate the same elements for
the $S$-matrix~\cite{Rebhan:1987cd}.
Therefore, one might argue that if the $S$-matrix contains all the
information about physical observables, then the Vilkovisky effective
action gives no new predictions. As shown in~\cite{KKT-88}, however, there are
physical quantities (e.g., critical temperatures) that in principle cannot
be directly obtained from scattering amplitudes and they depend on the choice of
gauge and parametrization of the quantum fields, if calculated using the standard
effective action. In this respect, the Vilkovisky effective action can give
unambiguous results which may be verified by experiments. The examples shown in
the main part of this work represent another application in which this formalism
can be useful.

Secondly, there are other ways of constructing covariant effective actions
besides~\eqref{UEA_def}. In particular, a generalization of the
unique effective action was presented by DeWitt~\cite{DeWitt-ea},
which is usually called ``Vilkovisky--DeWitt effective action''.
This more intricate definition coincides with~\eqref{UEA_def} in
the one-loop approximation, and has the advantage of yielding a
perturbative expansion in terms of one-particle irreducible
diagrams (which was a major problem in Vilkovisky's original
proposal)~\cite{Rebhan:1986wp,Rebhan:1987cd,Ellicott:1987ir}.
Since the considerations in the present work are based on the
one-loop renormalization group equations
(and in Sec.~\ref{s3} we argued why higher-loop effects are suppressed~\cite{UEA-RG}),
the definition~\eqref{UEA_def} is sufficient for our purposes, and
our results also hold within the more elaborate construction by DeWitt.

Last but not least, a common reservation about the use of the Vilkovisky--DeWitt
effective action when applied specifically to gravitational theories is because
the metric in the space of fields~\eq{sm} is not uniquely defined without the
prescription (iii). Notice that (iii) is not needed for the invariance of the Vilkovisky--DeWitt
effective action.
Indeed, by using just~\eq{sm} it is possible to construct a
whole family of effective actions parametrized by the arbitrary constant $a$,
each of them being gauge and parametrization invariant.
Vilkovisky's prescription
fixes $a$ in an uniquely way which only
depends on the gravitational theory in question.
Whether it is the correct prescription
is still an open question that might be decided only at the experimental level
(in this regard, see also~\cite{KKT-88,Vilkovisky:1992pb}).

\section*{Acknowledgments}
I.Sh. is grateful to A. Belyaev for useful discussions of the
particle physics aspects of the restrictions on the SSB.
The work of I.Sh. is partially supported by Conselho Nacional de
Desenvolvimento Cient\'{i}fico e Tecnol\'{o}gico - CNPq under the
grant 303635/2018-5, by Funda\c{c}\~{a}o de Amparo \`a Pesquisa
de Minas Gerais - FAPEMIG under the grant PPM-00604-18, and by
Ministry of Education of Russian Federation under the project
No. FEWF-2020-0003.



\begin{thebibliography}{99}

\bibitem{OUP} I.L. Buchbinder and I.L. Shapiro,
\textit{Introduction to Quantum Field Theory with Applications to
Quantum Gravity,} (Oxford University Press, 2021).

\bibitem{Peebles2002}
P.J.E.~Peebles and B.~Ratra,
\textit{The Cosmological Constant and Dark Energy,}
Rev. Mod. Phys. \textbf{75} (2003) 559, 
astro-ph/0207347.

\bibitem{Copeland2006}
E.J.~Copeland, M.~Sami and S.~Tsujikawa,
\textit{Dynamics of dark energy,}
Int. J. Mod. Phys. D \textbf{15} (2006)  1753, 
hep-th/0603057.

\bibitem{Sahni2002}
V.~Sahni, T.D.~Saini, A.A.~Starobinsky and U.~Alam,
\textit{Statefinder: A New geometrical diagnostic of dark energy,}
JETP Lett. \textbf{77} (2003) 201, 
astro-ph/0201498.

\bibitem{Weinberg89}  S. Weinberg,
{\it The cosmological constant problem,}
Rev. Mod. Phys. \textbf{61} (1989) 1.

\bibitem{CC-nova} I.L. Shapiro, J. Sol\`{a},
{\it Scaling behavior of the cosmological constant:
Interface between quantum field theory and cosmology,}
JHEP {\bf 02} (2002) 006,
hep-th/0012227.

\bibitem{Brown} L.S. Brown,
\textit{Quantum Field Theory,}
(Cambridge University Press, Cambridge,
1994).

\bibitem{nelspan82} B.L. Nelson and P. Panangaden,
{\it Scaling behavior of interacting quantum fields in
curved space-time,}
Phys. Rev. {\bf D25} (1982) 1019.

\bibitem{frts82} E.S. Fradkin and  A.A. Tseytlin,
{\it Renormalizable asymptotically free quantum theory of gravity,}
Nucl. Phys. {\bf B201} (1982) 469.

\bibitem{PoImpo} I.L.~Shapiro,
{\it Effective action of vacuum: semiclassical approach},
Class. Quant. Grav. {\bf 25} (2008) 103001,
arXiv:0801.0216.

\bibitem{Babic}
A. Babic, B. Guberina, R. Horvat, and H. Stefancic,
\textit{Renormalization group running of the cosmological constant
and its implication for the Higgs boson mass in the standard model,}
Phys. Rev. \textbf{D65} (2002) 085002,
hep-ph/0111207.

\bibitem{DCCrun} I.L.~Shapiro, J.~Sol\`{a},
{\it On the possible running of the cosmological 'constant'},
Phys. Lett. {\bf B682} (2009)  105,   hep-th/0910.4925.

\bibitem{Opher-2004} R.~Opher and A.~Pelinson,
\textit{Studying the decay of the vacuum energy with the observed
density fluctuation spectrum,}
Phys. Rev. D \textbf{70} (2004) 063529,
astro-ph/0405430.

\bibitem{Gruni} I.L.~Shapiro, J.~Sol\`{a} and H.~Stefancic,
\textit{Running G and Lambda at low energies from physics
at M(X): Possible cosmological and astrophysical implications,}
JCAP \textbf{01} (2005) 012,
hep-ph/0410095.

\bibitem{Babic-setting}
A. Babic, B. Guberina, R. Horvat, and H. Stefancic,
\textit{Renormalization-group running cosmologies.
A Scale-setting procedure,}
Phys. Rev. \textbf{D71} (2005) 124041,
astro-ph/0407572.

\bibitem{AC} T.~Appelquist and J.~Carazzone,
{\it Infrared Singularities and Massive Fields,}
Phys. Rev.  {\bf D11} (1975) 2856.

\bibitem{apco} E.V. Gorbar and I.L. Shapiro,
{\it Renormalization group and decoupling in curved space,}
JHEP {\bf 02} (2003) 021,
hep-ph/0210388;
{\it Renormalization group and decoupling in curved space:
II. The Standard Model and Beyond,}
JHEP {\bf 06} (2003) 004,
hep-ph/0303124.

\bibitem{CC-pert}
J.~Grande, J.~Sol\`{a}, J.C.~Fabris and I.L.~Shapiro,
\textit{Cosmic perturbations with running G and Lambda,}
Class. Quant. Grav. \textbf{27} (2010) 105004,
arXiv:1001.0259. 

\bibitem{Jhonny-GrCo}
J.A.~Agudelo Ruiz, J.C.~Fabris, A.M.~Velasquez-Toribio and
I.L.~Shapiro,
\textit{Constraints from observational data for a running
cosmological constant and warm dark matter with curvature,}
Grav. Cosmol. \textbf{26} (2020) 316, 
arXiv:2007.12636.

\bibitem{seesaw} A.~Accioly, B.L.~Giacchini and I.L.~Shapiro,
{\it On the gravitational seesaw in higher-derivative gravity},
Eur. Phys. J.  {\bf C77} (2017) 540,
gr-qc/1604.07348.

\bibitem{don-reviews} J.F.~Donoghue,
\textit{The effective field theory treatment of quantum gravity,}
AIP Conf. Proc. \textbf{1483} (2012) 
73, 
arXiv:1209.3511.

\bibitem{Burgess}  C.P.~Burgess,
{\it Quantum gravity in everyday life: General relativity as an
effective field theory,}
Living Rev. Rel. {\bf 7} (2004) 5,
gr-qc/0311082.

\bibitem{DalMaz} D.A.R.~Dalvit and F.D.~Mazzitelli,
\textit{Geodesics, gravitons and the gauge fixing problem,}
Phys. Rev. \textbf{D56} (1997) 7779, 
hep-th/9708102.

\bibitem{effQG} T. de Paula Netto, I.L. Shapiro, and L. Modesto,
\textit{Universal leading quantum correction to the Newton potential,}
Eur. Phys. J. \textbf{C82} (2022)  160,
arXiv:2110.14263. 

\bibitem{Vilk-Uni} G.A. Vilkovisky,
{\it The unique effective action in quantum field theory,}
Nucl. Phys. {\bf B234} (1984) 125. 

\bibitem{DeWitt-ea}
{B.S.~DeWitt},
\textit{The effective action},
in \textit{Quantum Field Theory and Quantum Statistics},
Essays in honor of the sixtieth birthday of {E.S.~Fradkin},
edited by C.J.~Isham, I.A.~Batalin and G.A.~Vilkovisky,
(Hilger, Bristol, 1987).


\bibitem{TV90} T.~Taylor and G.~Veneziano,
{\it Quantum gravity at large distances and the cosmological
constant},
Nucl. Phys. \textbf{B345} (1990) 210. 

\bibitem{UEA-contas}
B.L.~Giacchini, T.~de Paula Netto and I.~L.~Shapiro,
\textit{Vilkovisky unique effective action in quantum gravity,}
Phys. Rev.  \textbf{D102} (2020) 106006,
arXiv:2006.04217. 

\bibitem{UEA-RG} B.L.~Giacchini, T.~de Paula Netto and I.L.~Shapiro,
\textit{On the Vilkovisky-DeWitt approach and renormalization group
in effective quantum gravity,}
JHEP \textbf{10} (2020) 011,
arXiv:2009.04122.

\bibitem{KKT-88} K. Kobes, G. Kunstatter and D.J. Toms,
\textit{The Vilkovisky-DeWitt Effective Action:
Panacea or Placebo?},
 in \textit{TeV Physics: 12th Johns Hopkins Workshop on Current Problems in Particle Theory},
edited by G. Domokos and S. Kovesi-Domokos,
(World Scientific, Singapore, 1988).

\bibitem{Weinberg-87antrop}  S.~Weinberg,
{\it Anthropic bound on the cosmological constant,}
Phys. Rev. Lett.  {\bf 59} (1987) 2607.

\bibitem{aref} I.Ya. Arefeva, A.A. Slavnov and L.D. Faddeev,
{\it Generating functional for the $S$-matrix in gauge-invariant
theories},
Theor. Math. Phys. {\bf 21} (1975) 1165 \
[Version in Russian: Teor. Mat. Fiz. {\bf 21} (1974) 311].

\bibitem{volatyu} B.L. Voronov, P.M. Lavrov and I.V. Tyutin,
{\it Canonical transformations and the gauge dependence in
general gauge theories},
Sov. J. Nucl. Phys. {\bf 36} (1982) 498.

\bibitem{a} I.L. Shapiro and A.G. Jacksenaev,
{ \it Gauge dependence in higher derivative quantum
gravity and the conformal anomaly problem},
Phys. Lett. {\bf B324} (1994) 286.

\bibitem{JDG-QG}
J.D. Gon\c{c}alves, T. de Paula Netto and I.L. Shapiro,
{\it On the gauge and parametrization ambiguity in quantum gravity,}
Phys. Rev. {\bf D97} (2018) 026015,
arXiv:1712.03338.

\bibitem{bavi85}  A.O. Barvinsky and G.A. Vilkovisky,
\textit{The generalized Schwinger-DeWitt technique in gauge theories
and quantum gravity,}
Phys. Repts. {\bf 119} (1985) 1.

\bibitem{Huggins:1987zw}
S.R.~Huggins, G.~Kunstatter, H.P.~Leivo and D.J.~Toms,
\textit{The Vilkovisky-de Witt effective action for quantum gravity},
Nucl. Phys. \textbf{B301} (1988) 627.

\bibitem{Fradkin:1983nw}
  E.S.~Fradkin and A.A.~Tseytlin,
  {\it On the new definition of off-shell Effective Action},
  Nucl.\ Phys.\ {\bf B234} (1984) 509.

\bibitem{Rebhan:1987cd}
  A.~Rebhan,
  {\it Feynman Rules and $S$-Matrix Equivalence of the Vilkovisky-deWitt Effective Action},
  Nucl.\ Phys.\ {\bf B298} (1988) 726.

\bibitem{Ellicott:1987ir}
 P.~Ellicott and D.~Toms,
 {\it On the New Effective Action in Quantum Field Theory},
Nucl. Phys. {\bf B312} (1989) 700.


\bibitem{Abbot82}  L.F.~Abbott,
\textit{Introduction to the Background Field Method,}
Acta Phys. Polon. \textbf{B13} (1982) 33.

\bibitem{BDW-65} B.S.~DeWitt,
{\it Dynamical theory of groups and fields}
(Gordon and Breach, New York, 1965).

\bibitem{J.L.Synge:1960zz}  J.L.~Synge,
{\it Relativity: the general theory}
(North-Holland, Amsterdam, 1960).

\bibitem{Kunstatter:1986qa}
  G. Kunstatter,
  {\it Vilkovisky's Unique Effective Action: An introduction and explicit calculation},
  in {\it Super Field Theories},
  proceedings of NATO Advanced Research Workshop on Superfield Theories,
  edited by H.C.~Lee, V.~Elias, G.~Kunstatter, R.B.~Mann and K.S.~Viswanathan,
  NATO ASI Series B, Vol. 160 (Plenum, New York, 1987).

\bibitem{Rebhan:1986wp}   A.~Rebhan,
{\it The Vilkovisky-DeWitt Effective Action and its application to 
Yang-Mills Theories},
Nucl.\ Phys.\ {\bf B288} (1987) 832.

\bibitem{Vilkovisky:1992pb}
G.A.~Vilkovisky,
{\it Effective action in quantum gravity},
Class. Quant. Grav. \textbf{9} (1992) 895.


\end{thebibliography}
\end{document}